\documentclass[twocolumn,twocolappendix]{aastex63}
\usepackage{graphicx}	% Including figure files
\usepackage{amsmath}	% Advanced maths commands
\usepackage{amssymb}	% Extra maths symbols

\shorttitle{Robust solar flux density calibration at the MWA}
\shortauthors{Kansabanik et al.}
\begin{document}

\title{Robust absolute solar flux density calibration for the Murchison Widefield Array}

\correspondingauthor{Devojyoti Kansabanik}
\email{dkansabanik@ncra.tifr.res.in}

\author[0000-0001-8801-9635]{Devojyoti Kansabanik}
\affiliation{National Centre for Radio Astrophysics, Tata Institute of Fundamental Research, Pune University, Pune 411007, India}
%\nocollaboration

\author[0000-0002-2325-5298]{Surajit Mondal}
\affiliation{Center for Solar-Terrestrial Research, New Jersey Institute of Technology, 323 M L King Jr Boulevard, Newark, NJ 07102-1982, USA}
\affiliation{National Centre for Radio Astrophysics, Tata Institute of Fundamental Research, Pune University, Pune 411007, India}

\author[0000-0002-4768-9058]{Divya Oberoi}
\affiliation{National Centre for Radio Astrophysics, Tata Institute of Fundamental Research, Pune University, Pune 411007, India}

\author[0000-0002-1741-6286]{Ayan Biswas}
\affiliation{National Centre for Radio Astrophysics, Tata Institute of Fundamental Research, Pune University, Pune 411007, India}

\author[0000-0001-8496-252X]{Shilpi Bhunia}
\affiliation{Astronomy and Astrophysics Section, Dublin Institute for Advanced Studies, Dublin 2, D02 XF86, Ireland}
\affiliation{School of Physics, Trinity College Dublin, Dublin 2, Ireland}
\affiliation{Indian Institute of Science Education and Research, Pune, India}

\begin{abstract}
Sensitive radio instruments are optimized for observing faint astronomical sources, and usually need to attenuate the received signal when observing the Sun. There are only a handful of flux density calibrators that can comfortably be observed with the same attenuation setup as the Sun. Additionally, for wide field-of-view (FoV) instruments like the Murchison Widefield Array (MWA) calibrator observations are generally done when the Sun is below the horizon, to avoid the contamination from solar emissions. These considerations imply that the usual radio interferometric approach to flux density calibration is not applicable for solar imaging. A novel technique, relying on a good sky model and detailed characterization of the MWA hardware, was developed for solar flux density calibration for MWA. Though successful, this technique is not general enough to be extended to the data from the extended configuration of the MWA Phase II. Here, we present a robust flux density calibration method for solar observations with MWA independent of the array configuration. We use different approaches $-$ the serendipitous presence of strong sources; detection of numerous background sources using high dynamic range images in the FoV along with the Sun; and observations of strong flux density calibrators with and without the additional attenuation used for solar observations—to obtain the flux scaling parameters required for the flux density calibration. Using the present method, we have achieved an absolute flux density uncertainty $\sim10\%$ for solar observations even in the absence of dedicated calibrator observations.
\end{abstract}


\section{Introduction}\label{sec:intro}
The quiet Sun is just about the source with the highest flux density in the meter wavelength radio sky, and its flux density can increase by multiple orders of magnitude during the periods of active emissions. 
It is challenging to build sensitive radio instruments capable of providing a linear response spanning the entire range from the very faint astronomical sources of usual interest to the mega-Jansky solar bursts. Usually the low-noise amplifiers, the very first elements in the signal chain, have sufficient dynamic range, but that does not hold true for the downstream signal chain.
Solar observations, therefore, typically require the use of {\textit {attenuators}} early on in the signal path to bring down the signal levels sufficiently, so that they lie in the linear regime of the downstream signal chain. Since attenuators can bring down the signal level by multiple orders of magnitude, it becomes hard to observe other astronomical sources, including most of the so-called flux density calibrators, with this attenuation in place. Flux density calibrators are comparatively bright radio sources whose flux density and spectra are known accurately and are typically used for absolute flux density calibration \citep[e.g][]{perley2017}. The standard flux density calibrator observations are generally done without any additional attenuators. Transferring the flux density calibration determined using observations of standard flux density calibrators to the Sun, generally observed with attenuators, is not straightforward, and it requires a detailed and accurate characterization of the attenuators. It is very effort intensive to do this characterization, and in practice, it is rarely available. This is also the case with the Murchison Widefield Array \citep[MWA;][]{lonsdale2009, tingay2013}.

Solar absolute flux density calibration usually relies on the use of the few very bright sources whose flux density is large enough for them to be observed using the same attenuation setting as solar observations. This approach is followed at the Nan\c{c}ay Radio Heliograph \citep[NRH; e.g.][]{bonmartin1983,avignon1989}, the Gauribidanur Radio Heliograph \citep[GRH; e.g.][]{sundaram2004,sundaram2005} and the Low Frequency Array \citep[LOFAR; e.g.][]{breitling2015,kontar2017}. Conventionally, calibrator observations are scheduled to be observed adjacent to and/or interspersed with observations of the target source(s). This is done to minimize the impacts of any drifts in instrumental gains in the time between observations of the calibrator and target source(s). The wide field of view (FoV) and high primary beam sidelobes of the MWA imply that observations of flux density calibrators while the Sun is above the horizon can have significant contamination from the solar signal. Hence, at the MWA, the practice has been to observe the flux density calibrators before sunrise and after sunset, and to broaden the pool of suitable calibrators, without the use of any additional attenuation. Additionally, during the early periods of MWA solar observations (2013 July to 2014 April), no calibrator observations in the same spectral configuration as the solar observations are available, further complicating the task of absolute flux density calibration for these observations.

Taking advantage of the fact that the MWA is a very well-characterized instrument, including a well-modelled primary beam, and the availability of reliable sky models \citep[e.g.][]{haslam1982}, \citet{oberoi2017} implemented an innovative and computationally lean nonimaging technique for absolute solar flux density calibration. They estimated that the uncertainty in the absolute flux density estimates obtained using this technique generally lie in the range 10 $-$ 60\%. 
\citet{suresh2017} and \citet{sharma2018} used it successfully to get reliable flux density calibration for nonimaging studies.  \citet{mohan2017} presented a technique to transfer this to interferometric images, which has been used in multiple works \citep[e.g.][]{mondal2019, mohan2019a, mohan2019b, mondal2020, Sharma2020, mohan2021}. Though successful, this technique relied crucially on the availability of multiple very short baselines ($\lesssim10\lambda$) to obtain the total solar flux density and also to average over baseline-to-baseline fluctuations arising largely from manufacturing tolerances. In its Phase II, the MWA now has multiple configurations \citep{wayth2018}. The extended configuration of MWA Phase II, which offers the highest angular resolution and is more desirable from a solar imaging perspective, has few baselines short enough to meet the criterion imposed by \citet{oberoi2017}. An independent approach  has also been used in the past to calibrate flux density of the solar MWA data \citep[e.g.][etc.]{mccauley2017,rahman2019}. In this approach, the solar maps obtained using the MWA data are scaled such that the integrated solar flux density matches the prediction of data-driven models like FORWARD \citep{gibson2016}. Limitations of this technique have been discussed in \citet{Sharma2020} and mainly stem from the inadequacies of the model.

With improved imaging quality now available from instruments like the MWA, the need for a precise solar flux density calibration has also become evident. Applications requiring precise flux density calibration include: (1) modeling of the gyrosynchrotron emission from the Coronal Mass Ejection (CME) plasma to estimate physical parameters of CMEs \citep[e.g.][]{bastian2001, tun2013, carley2017, mondal2020}; and (2) attempts at detailed comparison of the radio observable, including flux densities, obtained from elaborate models implemented by tools like FORWARD \citep{gibson2016} with the radio maps from the MWA \citep{Sharma2020}. 

The aim of this work is to use multiple independent approaches to perform robust absolute solar flux density calibration across the MWA frequency range. The approaches explored here include -- (1) Using the serendipitous presence of a strong source of known flux density in the range of few hundreds of Jansky,  in the wide FoV of MWA during solar observations; (2) using the high-quality imaging delivered by the recently developed Automated Imaging Routine for Compact Array for Radio Sun \citep[AIRCARS;][D. Kansabanik et al. in preparation]{mondal2019} to detect numerous galactic and extragalactic background sources, the flux densities of which are known independently from the GLEAM survey\citep{Wayth2015,walker2017} to
be in the range from a few to a few tens of Jansky in the wide FoV
of the MWA during solar observation; and (3) observations of a strong flux density calibrator source with and without the additional attenuation. Using these approaches and a database of MWA calibration solutions that has recently been made available \citep{Sokolwski2020}, we develop a flux density calibration scheme useful for all solar observations with the MWA.

This paper is organized as follows: Section \ref{sec:obs} describes the observations and data analysis, and Section \ref{sec:results} presents the results, including a comparison of the solar imaging quality with the GLEAM survey. It is followed by a short Section \ref{sec:apply} describing the method to apply the flux scaling parameters to any solar observation, and finally, we provide our conclusions in Section \ref{sec:conclusion}.

\section{Observations and data analysis}\label{sec:obs}
We have analyzed four datasets that were recorded with the MWA on 2014 May 04, July 12, and September 28, and on 2020 June 20. These were observed under the proposal code G0002. 
We refer to these observations by their observation dates, 20140504, 20140712, 20140928, and 20200620, in the following text. The details of these observations are given in Table \ref{tab:obs_settings}. For 20140928, data corresponding to frequencies above 132 $\mathrm{MHz}$ were bad and had to be discarded. The spectral setting of ``picket fence" refers to a spectral configuration where the 30.72 $\mathrm{MHz}$ of observing bandwidth is distributed in 12 sub-bands each of width 2.56 $\mathrm{MHz}$, centered close to 80, 88, 96, 108, 120, 132, 145, 160, 196, 210, 218, and 240 $\mathrm{MHz}$. Each of these sub-bands is referred to as a `picket'. In the contiguous mode, we observe the band from $\sim$119--151 $\mathrm{MHz}$. All of the observations were done using channel widths of 40 kHz and a time resolution of 0.5 s.
The observations on 2014 May 04, 2014 September 28, and 2020 June 20 are solar observations, while Virgo-A observations on 2014 July 12 were done at night time.

\begin{table*}
\centering
    \begin{tabular}{|p{1.4cm}|p{2.7cm}|p{1.5cm}|p{1.8cm}|p{1.5cm}|p{0.8cm}|p{1.3cm}|p{1.7cm}|p{1.8cm}|}
    \hline
        Label & Date & Attenuator settings (dB) & Array \newline{configuration} & Spectral mode & Sun in FoV  & Source(s) used & Imaging \newline{bandwidth} (MHz)  &  Imaging \newline{integration} time (s)\\ \hline \hline 
        20140504 & 2014 May 04 & 10 & Phase I & Picket fence & Yes & Multiple faint sources &  2.56 &  120\\ \hline 
        20140712 & 2014 July 12 & 10 & Phase I & Picket fence & No & Virgo A & 2.56 &  0.5\\ \hline
        20140928 & 2014 September 28 & 14 & Phase I & Picket fence & Yes & Virgo A & 0.16 & 0.5\\ \hline
        20200620 & 2020 June 20 & 10 & Phase II \newline{extended} & Contiguous & Yes & Crab &  0.16 &  9\\ \hline
    \end{tabular}
    \caption{Details of the different observations used.}
    \label{tab:obs_settings}
\end{table*}

\subsection{Approach to solar calibration and imaging}\label{sec:aircars_overview}
Conventionally, for interferometric imaging, one needs to determine the complex gains of the individual antennas of the array. 
The process of determining the amplitude part of these gains using observations of sources whose flux density and spectra are known a priori, the so-called flux density calibrator sources, is referred to as flux density calibration.
The same observations are used to determine the phase part of the complex gains to make the array coherent.
The phase solutions are further refined using observations of compact sources in the vicinity of the field of interest, the so-called phase calibrators, and finally the process of self-calibration.

Making use of the compact and centrally condensed array  configuration of the MWA and the very high flux density of the Sun, one can devise robust algorithms to determine the phases of the antenna gains without requiring any  calibrator observations. 
A relative calibration, across both antennas and bandwidth, of the amplitudes of the antenna gains can also be easily done.
AIRCARS implements this primary idea with many refinements and optimizations for snapshot spectroscopic solar imaging with the MWA.
As AIRCARS does not rely on observations of flux density calibrators, it cannot provide images on an absolute flux density scale, and is limited to providing relative bandpass amplitude calibration every 1.28 $\mathrm{MHz}$ for the antenna gains.
This work provides a robust prescription for obtaining absolute flux density calibrated solar radio images for the MWA, even in absence of observations of suitable flux density calibrators and irrespective of the array configuration used.

\subsection{Data analysis for 20140928 and 20200620} \label{sec:strong_source}
\begin{figure*}
    \centering
    \includegraphics[trim={5.5cm 1.5cm 5.5cm 0cm},clip,scale=0.5]{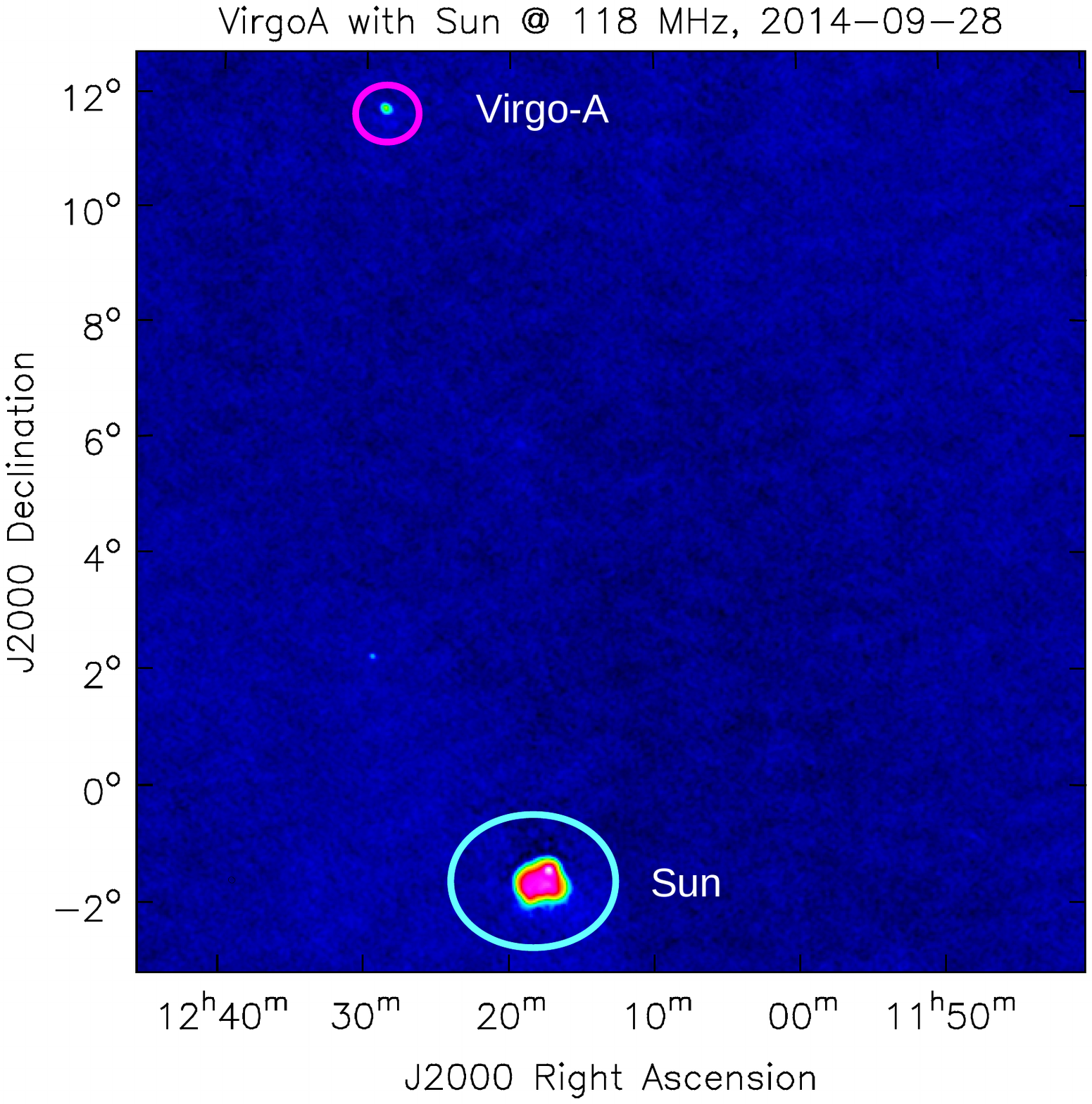}\includegraphics[trim={5.5cm 2cm 5.5cm 0cm},clip,scale=0.52]{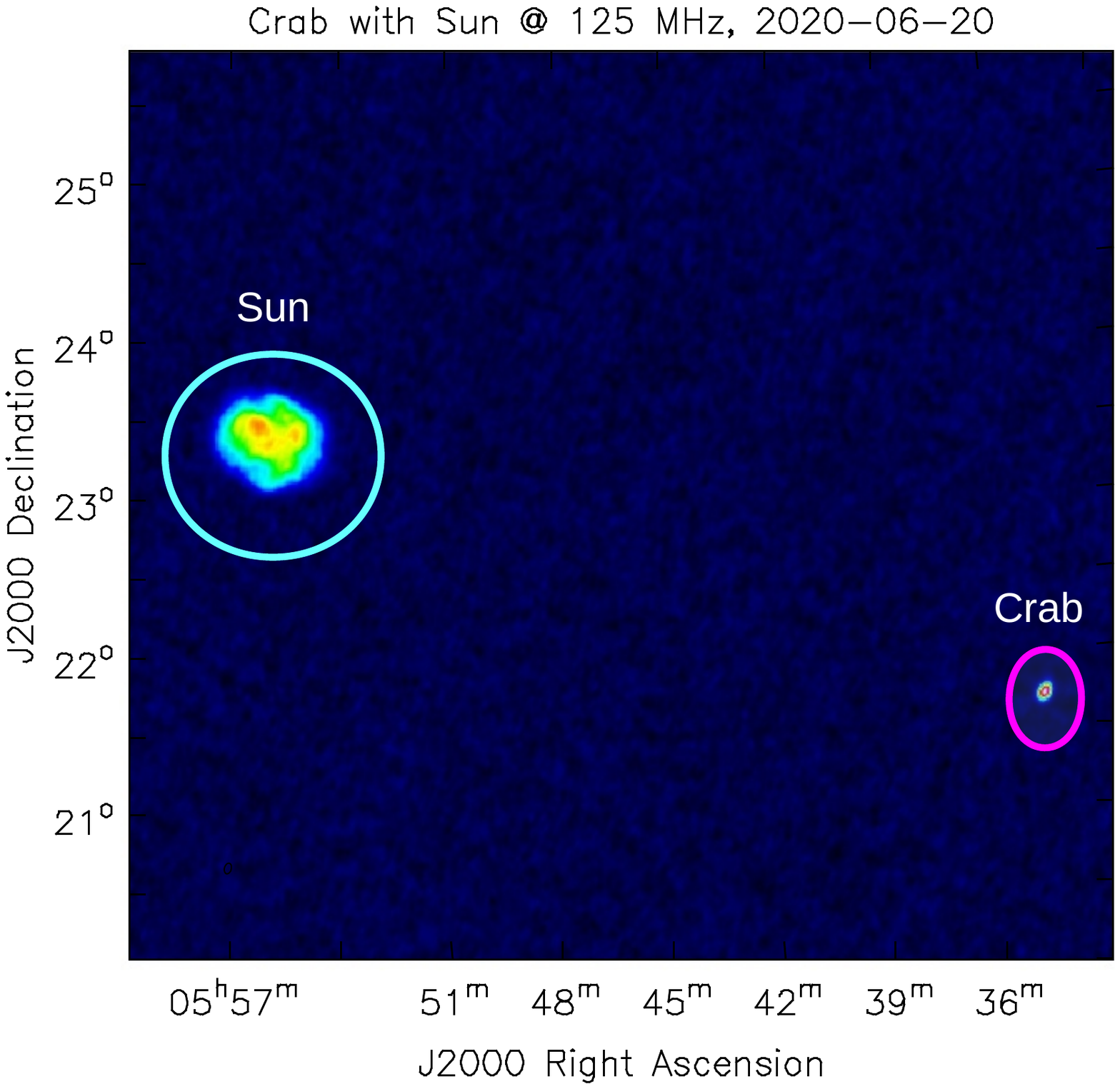}
    \caption{Left panel : A map at 118 $\mathrm{MHz}$ showing Virgo-A and the Sun. Right panel : A map at 125 $\mathrm{MHz}$ showing Crab and the Sun.}
    \label{fig:crab_and_virgo}
\end{figure*}

For these observations, the data analysis was fairly straightforward as the sources of interest, Crab and Virgo-A, are of comparable surface brightness to the quiet Sun \citep{perley2017}. The data were imaged using AIRCARS \citep{mondal2019}. The data quality for 20140928 is bad above 145 $\mathrm{MHz}$, and hence those data are not used here. While the spectral sampling of calibration scans were matched to solar observations, the calibrator sources were observed without the additional signal attenuation used for solar observations. Hence, for these observations, no fluxscale is derived from calibrator observations. An independent relative self-calibration is performed for each of the coarse channels to determine normalized gain solutions. Appropriate masks were used to ensure that both the Sun and Crab/Virgo-A were ``cleaned'' and included in the model used for self-calibration. The images of Virgo-A and Crab with the Sun in the FoV are shown in Figure \ref{fig:crab_and_virgo}. We used different temporal averaging for different datasets and listed in Table \ref{tab:obs_settings}. All other AIRCARS parameters were left at their default values. After the final images were obtained, the $\mathrm{imfit}$ task available in the Common Astronomical Software and Analysis \citep[\textsc{casa},][]{mcmullin2007} was used to estimate the flux density of these sources. Crab could be fitted well with a single Gaussian with residual flux density $<3$\%. For these two epochs we have reliable images for frequencies $<145$ $\mathrm{MHz}$. At these low-frequencies VirgoA can also be modeled with a small number of Gaussians components with a residual flux density $<4\%$. No significant emission feature is seen in either of the residual images. The flux-scaling factor for  reference epochs (Table \ref{tab:obs_settings}, $F_{ref}(\nu)={F_{cat}(\nu)}/{F_{app,ref}(\nu)}$, was computed by comparing the obtained primary beam corrected apparent integrated flux density, $F_{app,ref}(\nu)$, with the values, $F_{cat}(\nu)$, available from NASA/IPAC Extragalactic Database\footnote{https://ned.ipac.caltech.edu} or from the GLEAM catalog. 
The primary beam correction was obtained using the model by \citet{sokolowski17}. We calculate the uncertainty on $F_{ref}(\nu)$, $\Delta F_{ref}(\nu)$, by considering only the errors on the measured flux densities from the image. It is given by $\Delta F_{ref}(\nu)= F_{ref}(\nu) \times \Delta F_{intg}(\nu)/F_{intg}(\nu)$ where $F_{intg}(\nu)$ and $\Delta F_{intg}(\nu)$ are the integrated flux density of the source and the uncertainty on it, respectively.
We note that the noise in the final radio image follows a Gaussian distribution before a correction for the primary beam is applied.
However, quantities like $F_{ref}(\nu)$ and $\Delta F_{ref}(\nu)$ can only be computed after correcting for the primary beam.
Hence, the distribution of $\Delta F_{ref}(\nu)$ is not expected to follow a Gaussian as is discussed in Sec. \ref{sec:other_sources}.

\subsection{Data analysis for 20140504} \label{subsec:weak_sources}
As none of the sources in the FoV were strong enough to be detectable in snapshot images in the presence of the Sun and the solar attenuation, this dataset required significantly more involved analysis than the other three. Some of the requirements of this analysis were not met by AIRCARS in its original form. AIRCARS had focused primarily on the spectroscopic snapshot imaging use case.
As it was always working with only one spectral channel at a time, AIRCARS did not need relative bandpass calibration capability. Detecting the much fainter background sources, however, requires averaging over the entire 2.56 $\mathrm{MHz}$ bandwidth of the spectral pickets. Since the instrumental bandpass amplitude and phase is not constant over frequency, we require bandpass calibration before making the image over 2.56 $\mathrm{MHz}$.  The large difference between the observation times of calibrator sources and the Sun can potentially lead to changes in the shapes of the bandpass amplitudes and the phases between the calibrator and target sources. We perform independent {\em relative} bandpass calibration for each 2.56 $\mathrm{MHz}$ picket, but do not determine any {\em absolute} bandpass gain to account for the variation in spectral gain across the MWA band spanned by the pickets.

\subsubsection{Imaging the faint sources}
\begin{figure}
    \centering
    \includegraphics[scale=0.3]{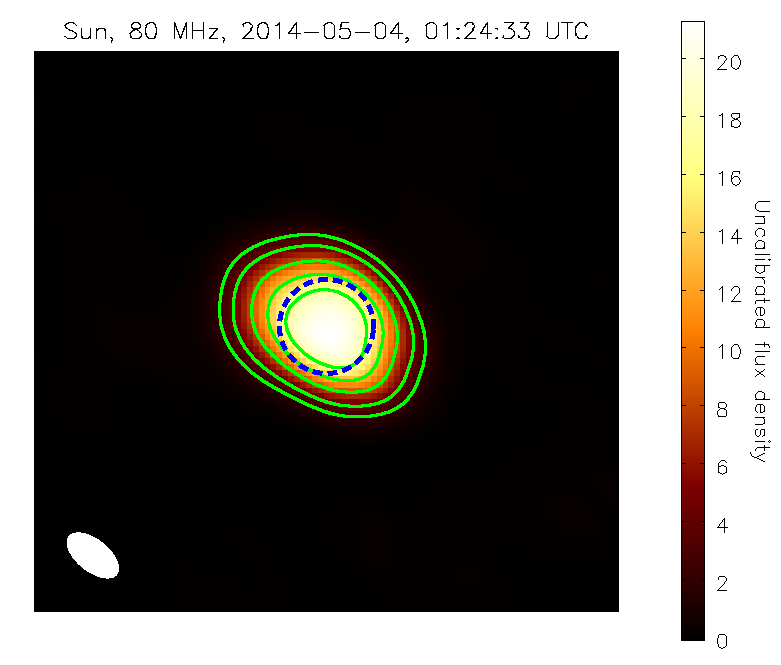}
    \caption{A map of the radio Sun at 80 $\mathrm{MHz}$.% at 2014-05-04 01:24:33 UTC is shown. 
    The green contours are at 0.1, 0.2, 0.4, 0.6, and 0.8 of the peak flux density. The blue dashed circle represents the optical disk of the Sun, and the point spread function is shown by the ellipse at the bottom left.}
    \label{fig:SUN}
\end{figure}

To overcome the limitations just mentioned AIRCARS was augmented to include bandpass self-calibration. This is a part of a larger effort to develop the next and more capable incarnation of AIRCARS, and will be presented in detail in a forthcoming paper (D. Kansabanik et al., in preparation). The bandpass self-calibration required us to find a way to deal with the degeneracy between the instrumental bandpass shape and the intrinsic spectral structure of the source (Sun). For convenience, we assume that the spectral structure of the Sun to be flat over the calibration bandwidth of 2.56 $\mathrm{MHz}$.
These data were obtained when the Sun was quiet (Fig. \ref{fig:SUN}).
The quiet Sun meterwave radio emission arises due to bremsstrahlung from the million Kelvin coronal plasma.
As the optical depth of this plasma approaches unity, a blackbody spectrum is an acceptable approximation.
The blackbody spectrum departs from a flat one by $\lesssim5\%$  over a fractional bandwidth of $\lesssim2.5\%$.
Hence, ignoring the spectral variation over our narrow bandwidths of observation is a reasonable assumption.

The basic approach was to image and model the solar emission, subtract it from the observed visibilities, and image the residual visibilities to look for the background sources. 
This was implemented as follows: augmented AIRCARS was used to generate a solar image integrated over the full 2.56 $\mathrm{MHz}$ bandwidth of each spectral picket at 10 $\mathrm{s}$ resolution.
This solar emission was modeled using the \textsc{CASA} \textit{tclean} task without the \textit{w-projection} algorithm switched on. A mask that limited the cleaning only to the solar disk was used. The final solar images and clean models were generated and the model visibilities  for the Sun corresponding to the clean model were subtracted from the calibrated visibilities using the \textsc{CASA} \textit{uvsub} task. The residual visibilities thus obtained over the 2 minutes were imaged using the \textsc{CASA} \textit{tclean} task with the \textit{w-projection} algorithm switched on and with uniform weighting. While imaging the residual visibilities, the phase center was shifted close to the direction of the peak of the primary beam.

\subsection{Data analysis for 20140712}
These observations were designed to look at a strong calibrator  with  and  without  the  attenuation  usually  used  for  solar  observations, and were done at night. For this observation, since the Sun was not present in the field of view, the analysis was straightforward. We follow a similar calibration approach what was implemented AIRCARS. Each of the 12 bands of width 2.56 $\mathrm{MHz}$ was calibrated using a normalized bandpass of the each sub-band using the $\textit{bandpass}$ task of CASA \citep[][]{mcmullin2007} for every 10 $\mathrm{s}$ time interval. This  imaging was done for each of these 12 bands. At higher frequencies, Virgo-A is resolved by the MWA baselines. Hence, the ``Python Blob Detector and Source Finder"  \citep[PyBDSF;][]{mohan2015} was used to fit multiple Gaussians to it, and the sum of the flux densities of all the Gaussian components was regarded as the total flux density of Virgo-A. 
We used the Virgo-A flux densities from \cite{perley2017} to obtain $F_{ref}(\nu)$ using the primary beam corrected values obtained from the 2.56 $\mathrm{MHz}$ images.

\section{Results}\label{sec:results}
\subsection{Determining $F_{ref}$ from weak sources} \label{sec:other_sources}
\begin{figure*}
\centering
\includegraphics[trim={11cm 11.5cm 11.5cm 2.5cm},clip,scale=2.5]{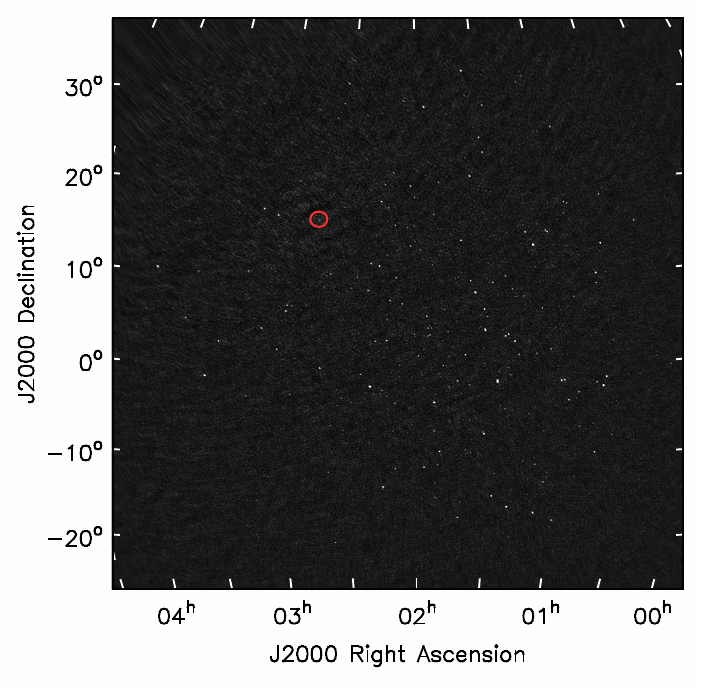}
\caption{Image centered at 80 $\mathrm{MHz}$, $\sim 60^{\circ}$ on each side, obtained over 2 minutes and 2.28 $\mathrm{MHz}$ after subtraction of modeled solar visibilities from the data. The image is from the observation on 2014 May 04, and the red circle with radius 2 $R_\odot$ is the region where the Sun was present.}
\label{fig:other_sources}
\end{figure*}

\begin{figure*}
    \centering
    \includegraphics[scale=0.8]{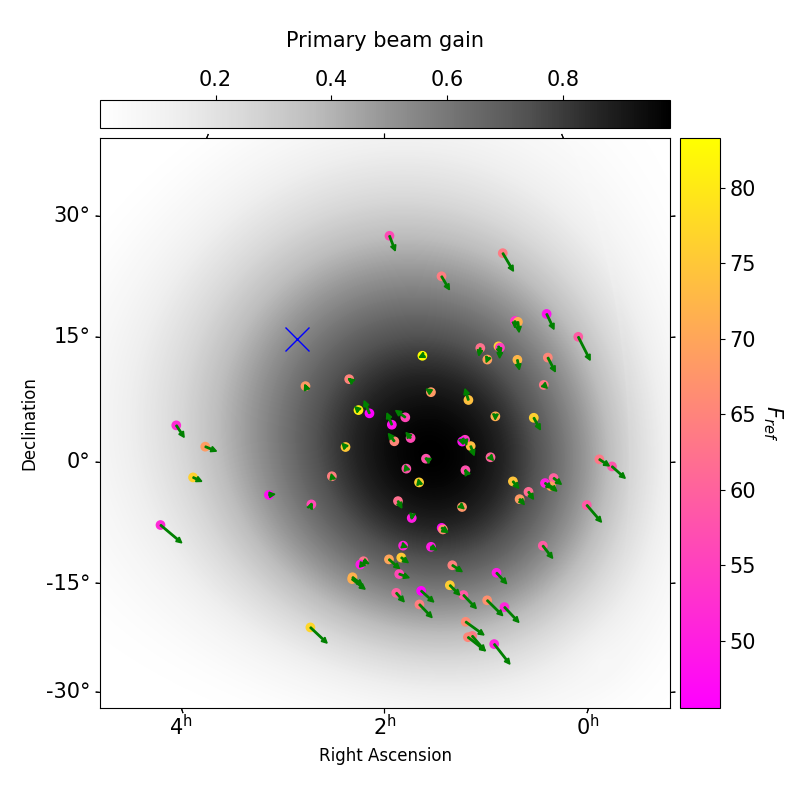}
    \caption{The locations of the subset of detected sources from the observation on 2014 May 04 used for determining $F_{ref}(\nu)$ are shown by colored dots. The colors of the dots denote their flux-scaling factor determined for each of the sources, and the primary beam gain at those locations is shown by the background gray scale. The arrows mark the shift of these source from their GLEAM catalog positions, multiplied by a factor of 15 to make them visible. The location of the Sun is shown by blue cross mark.}
    \label{fig:sources_detected_shited}
\end{figure*}

A prerequisite for determining $F_{ref}(\nu)$ is to either have one or more strong flux density calibrator sources in the FoV or several weak sources so that the fluctuations due uncertainty on individual estimates of $F_{ref}(\nu)$ can get averaged out. Naturally, detecting weak sources is more challenging, and to the best of our knowledge, imaging of multiple background sources in the vicinity of the Sun is yet to be demonstrated at meter wavelengths. Figure \ref{fig:other_sources} is perhaps the first image to show detection of numerous background sources with high signal-to-noise at low radio frequencies. The closest source to the Sun is at $\sim$20 $R_{\odot}$ with a flux density 4.9 $\mathrm{Jy}$. The $F_{ref}(\nu)$ obtained for different sources from 20140504 observation is shown in Figure \ref{fig:sources_detected_shited}. No systematic variation of $F_{ref}(\nu)$ with primary beam is apparent. All results stated here come from the observations on 20140504, and unless otherwise mentioned, they correspond to 80 $\mathrm{MHz}$. Dynamic range of the images at other frequencies below 145 $\mathrm{MHz}$ are comparable. 

Images at all frequencies were searched for sources independently before primary beam correction using the source finding software PyBDSF \citep{mohan2015}. PyBDSF was tuned such that only sources with at least a 7$\sigma$ significance, where $\sigma$ is the local rms in the image  as calculated by PyBDSF, were selected. The local rms is expected to drop with increasing angular distance from the Sun and was found to vary by a factor of about 2 across the FoV. It was also ensured that none of the sources included any pixels with flux density below 5$\sigma$. No sources were detected at frequencies above 145 $\mathrm{MHz}$. We attribute this to the combined effects of the flux density of the sources dropping at higher frequencies due to their typical negative spectral indices ($S_{\nu} \propto \nu^{\alpha}$) and the increasing solar flux density leading to higher system temperature (and lower sensitivity) at higher frequencies. In addition, the area of the FoV also decreases with increasing frequency. Additionally, the data quality was much poorer above 145 $\mathrm{MHz}$. The sensitivity achieved here fell short of what was needed for detection of sources with sufficient signal-to-noise at higher frequencies. 

All of the detected sources were carefully examined visually.
To avoid possible impacts of differences in the sensitivities of the images on extended low surface brightness features on $F_{ref}(\nu)$ estimates, we use only unresolved sources.
The detected compact sources were cross-matched with the MWA GLEAM survey catalog \citep{walker2017}. Cross-matching was done using {\it Aladin} \citep{bonnarel1999}, assuming that the sources can shift at most by $20^{'}$ from their reference positions.
The choice of maximum allowed shift was motivated by visual inspection of the sources detected in our image and the GLEAM sources. This shift can occur due to multiple reasons, including the following:

\begin{enumerate}
    \item The calibrator observations were carried out at night, whereas these observations come from close to solar transit. The refractive shift estimated and corrected for by the calibrator many hours apart and in a different direction are not applicable for solar observation.
    In addition, the process of self-calibration can also introduce an artificial direction-independent shift in the source locations. Both these effects contribute to giving rise to a  direction-independent shift in the source locations. 
    \item 
    Nighttime direction-dependent refractive shifts due to ionospheric structures have already been convincingly demonstrated \citep[e.g.][etc.]{Loi2015a,Loi2015b,Jordan2017,Walker2018,Helmboldt2020}.
    The daytime ionosphere can have about an order of magnitude higher electron column density.
    It is hence reasonable to expect significant direction-dependent shifts due to ionospheric refraction.
\end{enumerate}

Cross-matching catalogs that have been observed with different resolutions can be tricky. The GLEAM survey data used here were obtained with the MWA Phase I by observing the fields close to transit. In case of 20140504, the solar elevation was $31.{\circ}.3$ and hence the angular resolution is poorer than that of GLEAM. Conventionally, under such circumstances, only the sources that are unresolved in both images are used. To avoid losing the sources unresolved in our images but resolved in GLEAM, we pursue an approach inspired by \citet{rogers2004}. Sources which produce a single match in the GLEAM catalog and remain unresolved in GLEAM pose no challenge. Instances when a single unresolved source in our images matches more than one sources in GLEAM catalog require some thought. In such instances, we regard the integral of flux densities inside a region of the size of the point spread function (PSF) of our image centered on the brightest of the matched GLEAM sources as the effective flux density of the source. We have verified that the results obtained following this procedure and those obtained using unique unresolved GLEAM sources are consistent.

The GLEAM flux density at our frequency of interest is obtained by linear interpolation of flux densities measured by GLEAM at the frequency bands straddling the frequency of interest. Within the FWHM of the primary beam, we can detect all sources above a GLEAM flux density threshold of 10 $\mathrm{Jy}$ at 80 $\mathrm{MHz}$. The sensitivity levels are comparable at other frequencies. At locations where the primary beam gain is higher, we can detect much fainter sources. The weakest detected source had a GLEAM flux density of 4.6 $\mathrm{Jy}$ at 80 $\mathrm{MHz}$ near the peak of the primary beam. As we use a 7$\sigma$ threshold for choosing the sources, this implies that, close to the peak of the primary beam, the image rms is approximately 0.6 $\mathrm{Jy}$. 

Another issue to remain mindful of is the variability of flux densities \citep{Bell2013,Row2016,Lynch2017}. In order to avoid incurring errors in $F_{ref}(\nu)$ estimation by using sources with time variable flux densities, only sources for which the estimated $F_{ref}(\nu)$ lies between the 10$^{th}$ and 90$^{th}$ percentile are used. This criteria is met by 81 of the detected sources, and their locations are shown in Figure \ref{fig:sources_detected_shited}. The $F_{ref}(\nu)$ determined from each of these sources, the primary beam gain at their locations, and the observed shifts from their GLEAM catalog positions are also shown in Figure \ref{fig:sources_detected_shited}. The vectors showing the observed shifts of neighboring sources tend to be similar and vary smoothly across the large FoV. This systematic variation across the image is consistent with an ionospheric origin \citep{Loi2015a,Loi2015b,Jordan2017,Walker2018,Helmboldt2020}.

\begin{figure*}
    \centering
    \includegraphics[trim={0.5cm 0cm 0cm 0cm},clip,scale=0.405]{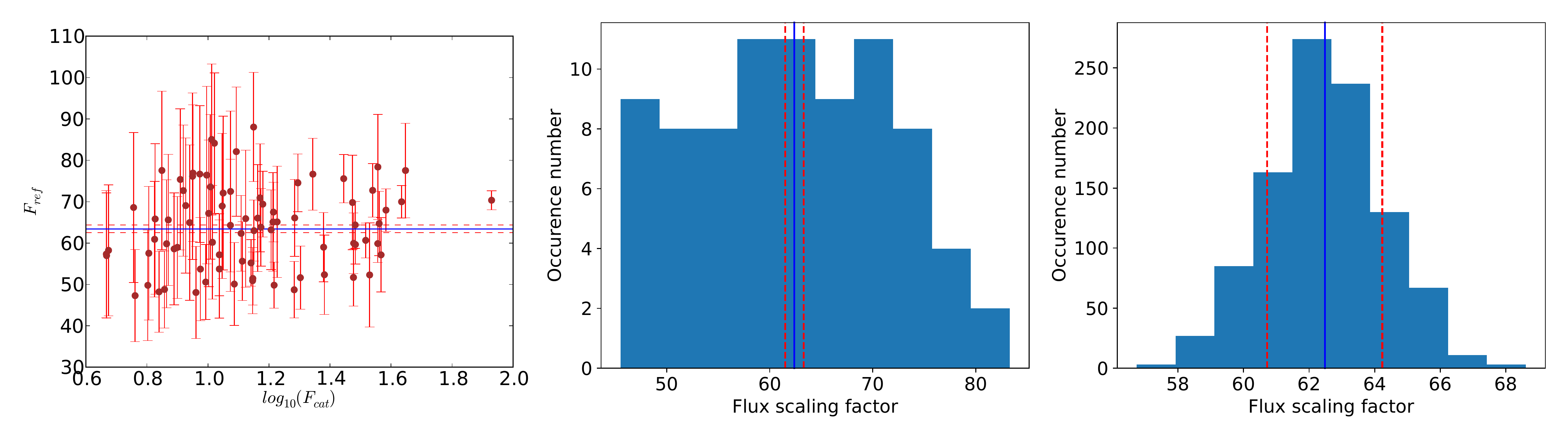}
    \caption{Left panel: Estimated $F_{ref}(\nu)$ against the logarithm of GLEAM flux density, $log_{10}(F_{cat})$, for individual sources at 80 $\mathrm{MHz}$. 
    Middle panel: Histogram of reference flux-scaling factors, $F_{ref}(\nu)$. Left panel: Histogram of $F_{ref}(\nu)$ obtained from random sampling. The blue solid line shows the median scaling factor, and the red dashed lines show the 1$\sigma$ uncertainty.}
    \label{fig:attenuator_gain_comparison}
\end{figure*}

\begin{figure*}
    \centering
    \includegraphics[trim={1.5cm 0.5cm 2cm 0.5cm},clip,scale=0.35]{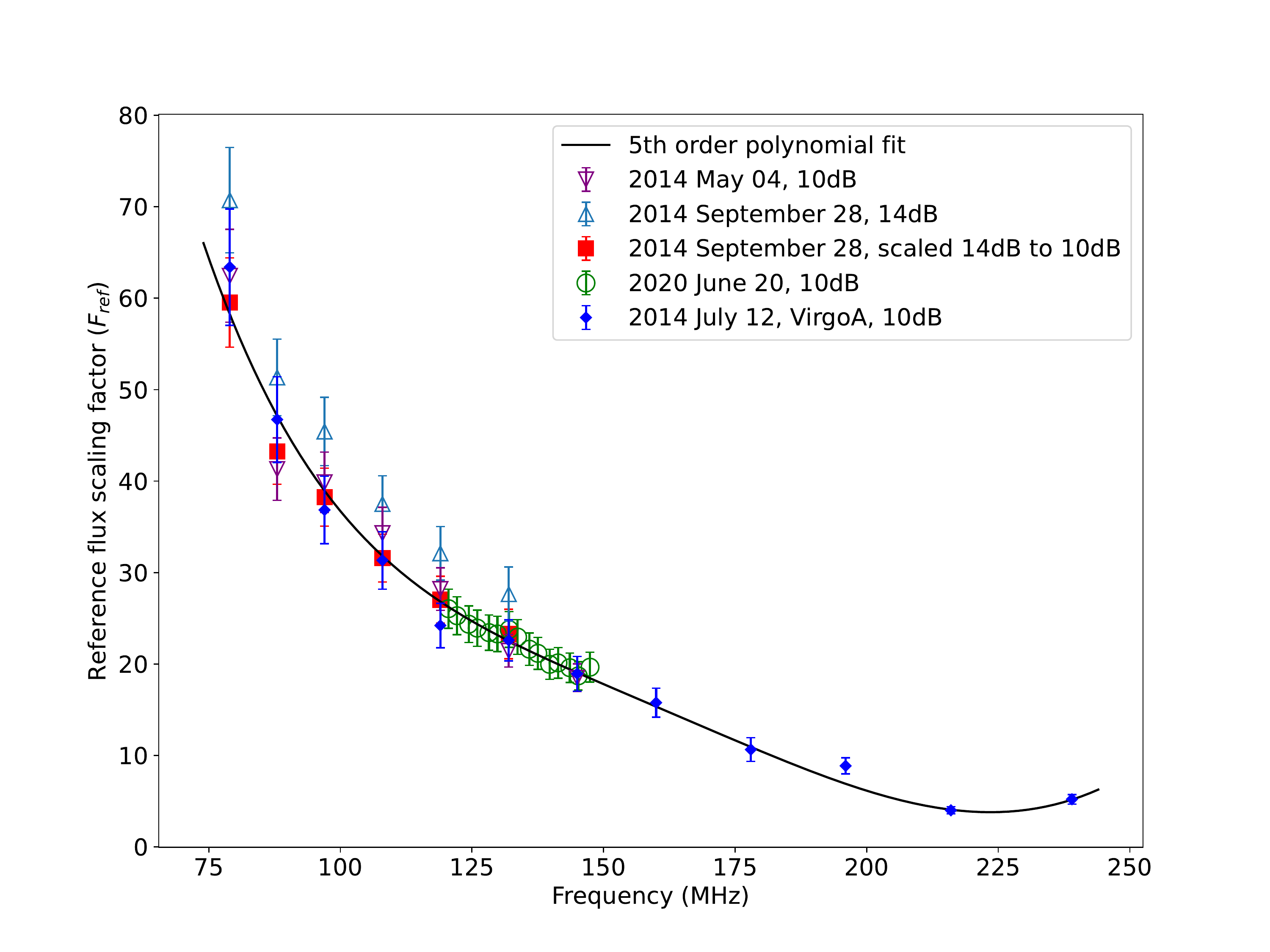}\includegraphics[trim={0.50cm 0 1cm 0},clip,scale=0.64]{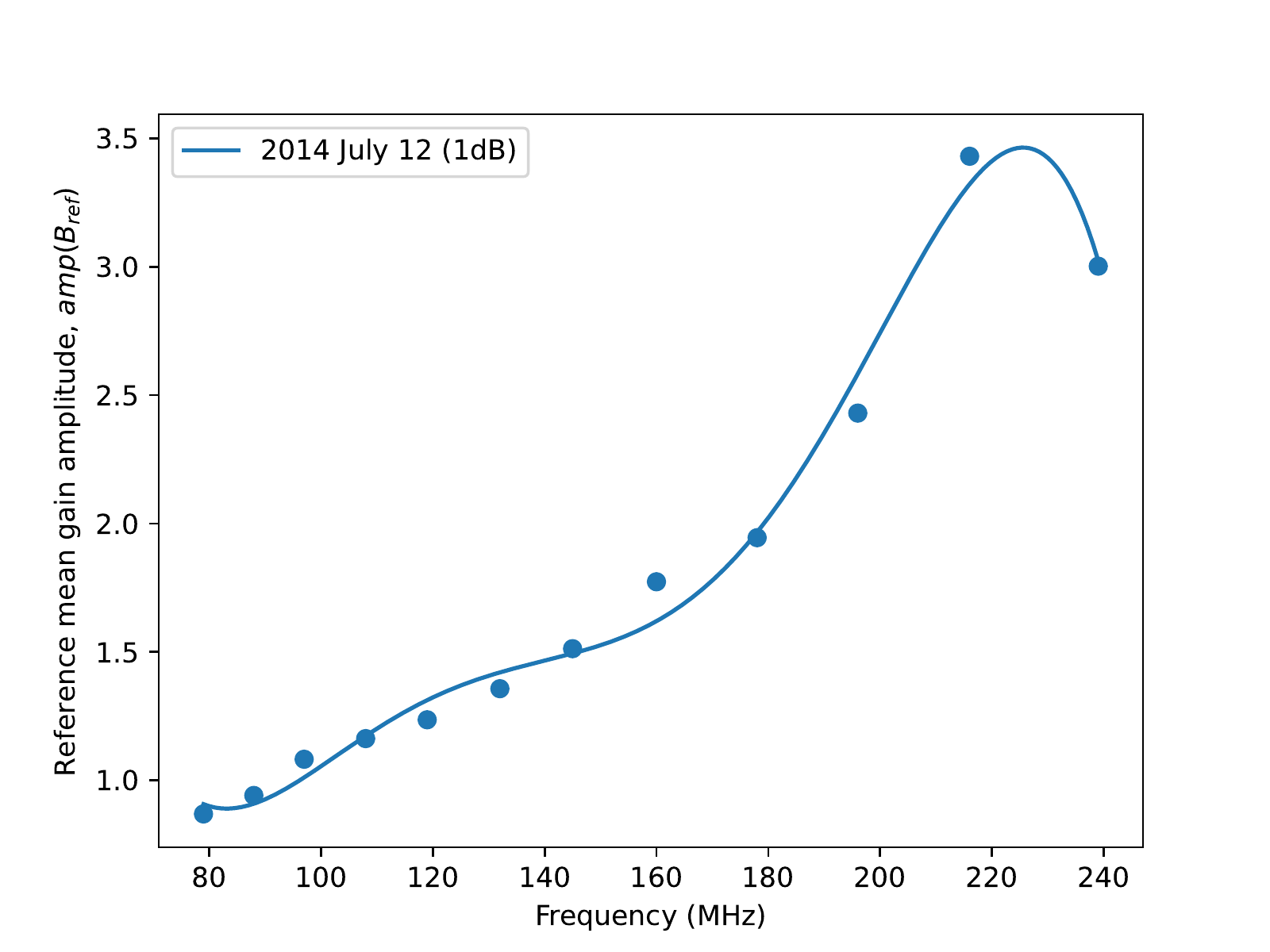} 
    \caption{Left panel : The reference flux scaling factor, $F_{ref}(\nu)$,  obtained for different observations. The black line shows the fitted polynomial. Right panel : Mean reference bandpass amplitude, $amp\ (B_{ref})$, for 2014 July 12.}
    \label{fig:fluxscale_polyfit}
\end{figure*}

\begin{figure}
    \centering
    \includegraphics[trim={0.40cm 0 1cm 0},clip,scale=0.56]{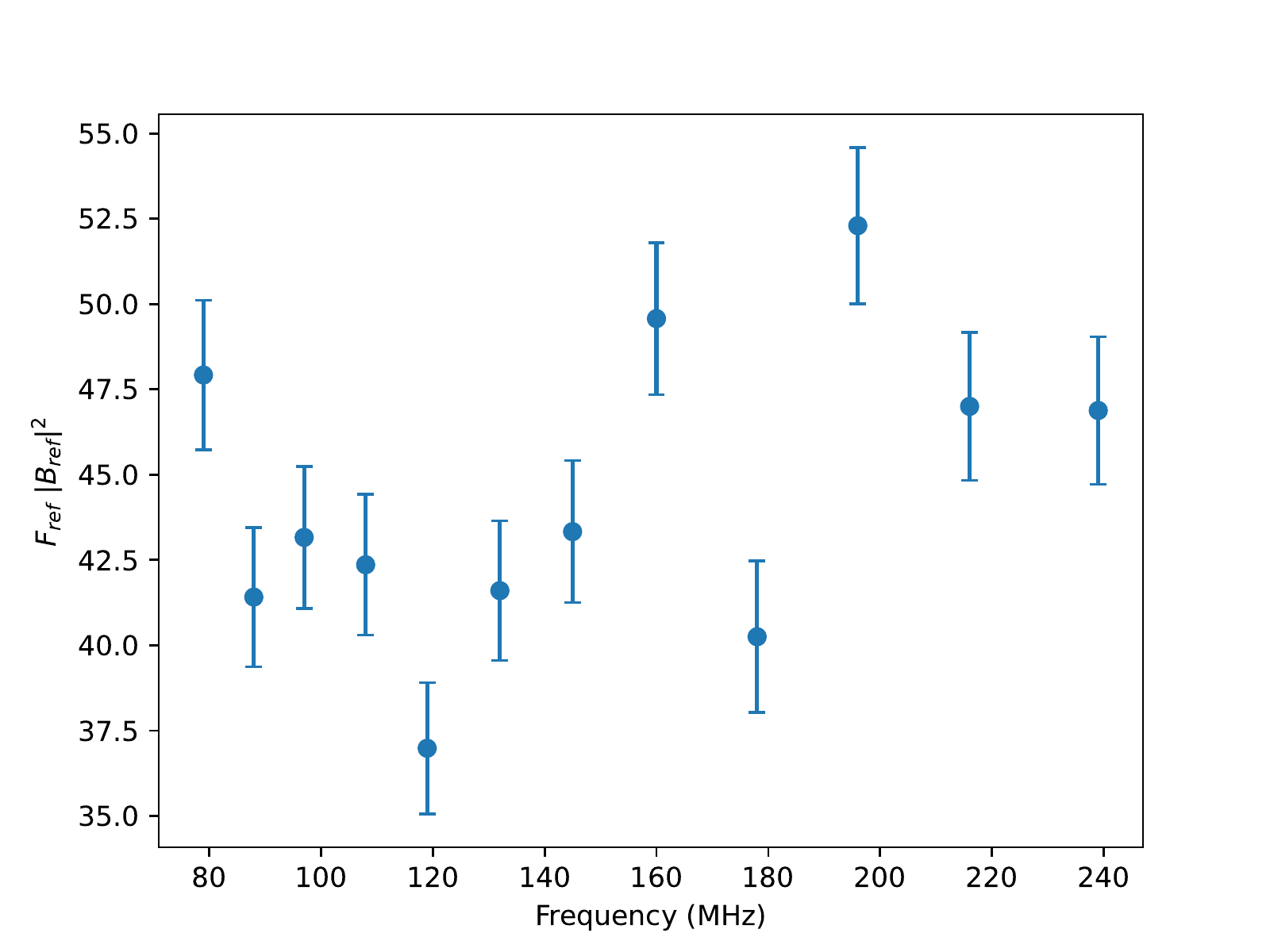} 
    \caption{Variation of the product of $F_{ref}(\nu)$ and $|B_{ref}(\nu)|^2$ with frequency.}
    \label{fig:fscale_mult_bref}
\end{figure}

\begin{figure}
    \centering
    \includegraphics[trim={0.5cm 0 1cm 1.4cm},clip,scale=0.45]{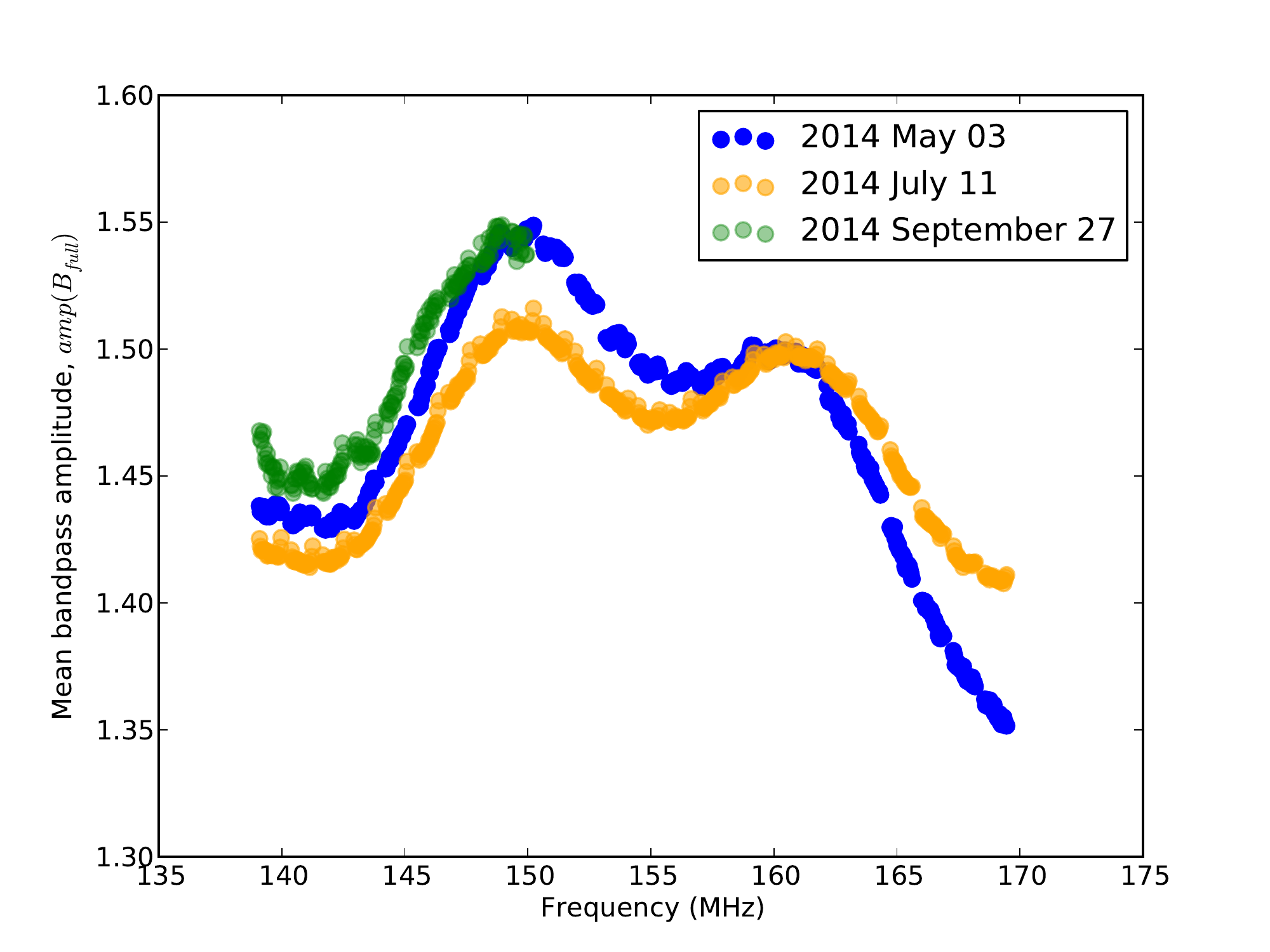} 
    \caption{Mean bandpass amplitudes, $amp\ (B_{full})$, for the calibrator observations. Three colors represent three observing epochs.}
    \label{fig:epochs_meanband}
\end{figure}
The left panel of Figure \ref{fig:attenuator_gain_comparison} shows $F_{ref}(\nu)$ estimated from individual sources, along with the associated uncertainty, as a function of their GLEAM flux densities. Since the levels of rms noise in the GLEAM images are much smaller than that of our image (Appendix \ref{app:1}), the contribution of error in GLEAM flux densities to $\Delta F_{ref}(\nu)$ is small and has been ignored. $\Delta F_{ref}(\nu)$ is calculated as mentioned in Section \ref{sec:strong_source}.

The middle panel of Fig. \ref{fig:attenuator_gain_comparison} shows the histogram of $F_{ref}(\nu)$.
The observed spread in the histogram results from the intrinsic uncertainties in the values of $F_{ref}(\nu)$ determined for each of the sources.
Using all sources at 80 $\mathrm{MHz}$, the inverse variance weighted mean of $F_{ref}(\nu)$ is found to be $62.4\pm0.9$.

As discussed in Sec. \ref{sec:strong_source}, the distribution of $\Delta F_{ref}$ is not expected to follow a Gaussian.  
The noise  characteristics  of  the  image  itself  were  Gaussian  before  the  primary  beam  correction, so it  is reasonable to expect the estimated uncertainty on each of the individual values of $F_{ref}$ determined to be drawn from their own Gaussian distribution -- one with a mean and sigma corresponding to $F_{ref}(\nu)$ and $\Delta F_{ref}(\nu)$ respectively.  
As a consistency check, 1000 sets were created, each with a value drawn from a Gaussian distribution with a mean and rms equal to $F_{intg}(\nu)$ and $\Delta F_{intg}(\nu)$, corresponding to each of the sources used. The
mean $F_{ref}(\nu)$ was then computed for each of these 1000 sets. The right panel of Fig. \ref{fig:attenuator_gain_comparison} shows the histogram of the mean $F_{ref}(\nu)$ from these 1000 sets, which shows a well-defined Gaussian. 
This approach yields an $F_{ref}(\nu)$ of $62\pm 2$ at 80 $\mathrm{MHz}$. This exercise makes it evident that there are no systematic errors associated with the determination of $F_{ref}(\nu)$, which otherwise would show as a departure from Gaussian distribution. This also verifies that the inverse variance weighted mean of the $F_{ref}$ shown in the left panel is consistent with the estimate from this exercise.

\subsection{Flux scale parameters}\label{sec:fluxscale_param}
In this section, we discuss about the origin of different flux density scaling parameters and how they are derived.
\subsubsection{Formulating flux scale parameters}\label{sec:formulating_fluxscale}
It is common practice in radio interferometric calibration to decompose the {\em antenna-dependent} instrumental bandpass gain, $B_{tot}(t,\ \nu)$ into a purely time-dependent part, $G_{mean}(t)$, and, a purely frequency-dependent part, $B_{full}(\nu)$. Following this approach, we decompose, $B_{tot}(\nu, t)$ 
as follows: 
\begin{equation}
\begin{split}
    B_{tot}(\nu,\ t) &= G_{mean}(t)\ B_{full}(\nu).
\end{split}
\label{eq:1}
\end{equation}
In the case of solar observations with the MWA, the conventional approach of flux density calibration is not followed, as has been discussed in Section \ref{sec:aircars_overview}. 
This section provides the prescription to implement  Eq. \ref{eq:1} for the MWA solar observation.

AIRCARS performs spectrally local bandpass calibration, normalized to unity, for each 2.56 $\mathrm{MHz}$ picket.
The MWA has a significant variation in its response across the band, which is not taken into account by the bandpass calibration done by AIRCARS.
To take the bandpass gain variation across pickets into account, we decompose $B_{full}(\nu)$ into the picket bandpass, $B_{picket}(\nu)$, and the inter-picket bandpass, $B_{inter}(\nu)$, both normalized to unity:
\begin{equation}\label{eq:bfull_vs_binter}
    B_{full}(\nu)=B_{picket}(\nu)\ B_{inter}(\nu)
\end{equation}
Equation \ref{eq:1} then can then be expressed as:
\begin{equation}
    \begin{split}
         B_{tot}(\nu, t)&= G_{mean}(t)\ B_{picket}(\nu)\ B_{inter}(\nu).\\
    \end{split}
    \label{eq:2}
\end{equation}
Of these, $B_{picket}(\nu)$ is already applied by AIRCARS. 
AIRCARS computes and corrects for $B_{picket}(\nu)$ when we use imaging over a picket of 2.56 $\mathrm{MHz}$. 
When imaging over much narrower bandwidth of 160 kHz, we simply assume $B_{picket}(\nu)=1$, as the bandpass is essentially flat over this tiny band. 
Only the two remaining terms need to be computed and corrected for. Thus, we can express $B_{tot}(\nu,\ t)$ as,
\begin{equation}\label{eq:btot_vs_gmean_and_binter}
B_{tot}(\nu,\ t)=G_{mean}(t)\ B_{inter}(\nu)
\end{equation}
To determine these parameters, we use the catalog flux densities, $F_{cat}(\nu)$, of the background galactic and extragalactic sources.
Before comparing the observed flux densities with $F_{cat}(\nu)$, the images are corrected for the primary beam response.
The flux densities observed in these primary beam corrected images are referred to as $F_{app}(\nu, t)$, the apparent flux density, which can differ from epoch to epoch.
In addition, for solar observations, we also need to account for the effect of the attenuators, $A(\nu,t)$, which in the most general case will be a function of time and frequency. 
$F_{app}(\nu,t)$ can then be expressed in terms of $F_{cat}(\nu)$, $B_{tot}(\nu,t)$ and $A(\nu,t)$ using Eqs. \ref{eq:1} and \ref{eq:btot_vs_gmean_and_binter} as:
\begin{equation}
\begin{split}
      F_{app}(\nu,t) & =A(\nu,t)\ |B_{tot}(\nu,t)|^2\ F_{cat}(\nu)\\
      &= A(\nu,t)\ |G_{mean}(t)|^2\ |B_{full}(\nu)|^2\ F_{cat}(\nu)\\
      &= A(\nu,t)\ |G_{mean}(t)|^2\ |B_{inter}(\nu)|^2\ F_{cat}(\nu)\\
      \frac{F_{cat}(\nu)}{F_{app}(\nu,\ t)}&=\frac{1}{A(\nu,\ t)\ |G_{mean}(t)|^2\ |B_{inter}(\nu)|^2}\\
      F_{scale}(\nu,\ t)&=\frac{1}{A(\nu,\ t)\ |G_{mean}(t)|^2\ |B_{inter}(\nu)|^2}\\
      F_{scale}(\nu,\ t)&=\frac{1}{A(\nu,\ t)\ |B_{tot}(\nu, t)|^2}\\
\end{split}
\label{eq:3}
\end{equation}

We regard observations on all four dates, 20140504,
20140712, 20140928, and 20200620, as the reference epoch,
$t_\mathrm{ref}$.
The justification for this is presented in Sec. \ref{sec:time-variation-atten}. For $t=t_{ref}$, $F_{scale}(\nu,\ t)$ is considered as $F_{ref}(\nu)$, which is similar to that defined in Sec. \ref{sec:strong_source} as follows:
\begin{equation}\label{eq:fref}
    \begin{split}
        F_{ref}(\nu)&=F_{scale}(\nu, t=t_{ref})\\
        F_{ref}(\nu)&=\frac{1}{A_{ref}(\nu)\ |G_{mean,ref}|^2\ |B_{inter}(\nu)|^2}\\
        F_{ref}(\nu)&=\frac{1}{A_{ref}(\nu)\ |B_{ref}(\nu)|^2}\\
        F_{ref}(\nu)&=\frac{F_{cat}(\nu)}{F_{app,ref}(\nu)}
    \end{split}
\end{equation}
where $B_{ref}(\nu)=G_{mean,ref}B_{inter}(\nu)$ is the un-normalized inter-picket bandpass gain without attenuation, $G_{mean,ref}$ is the $G_{mean}(t)$ for $t=t_{ref}$, and $F_{app,ref}(\nu)$ represents the apparent flux densities of the background sources for $t_{ref}$. The time variations in $F_{scale}(\nu, t)$ arise due to those in $G_{mean}(t)$. These time variations are computed by comparing the instrumental bandpass gain amplitudes for any epoch, $t$, with the overlapping frequency of the $B_{ref}(\nu)$. For most MWA solar observations, the calibrator observations were done without the use of attenuators. When applying the bandpass calibration to solar data, we also need to take the additional attenuation into account, and that is done using $F_{scale}(\nu, t)$.

\subsubsection{Computing flux scale polynomials}\label{sec:fluxscale_compute}
In this section, we discuss the recipe for computing $F_{ref}(\nu)$ and $B_{ref}(\nu)$. The left panel of Fig. \ref{fig:fluxscale_polyfit} shows the variation of $F_{ref}(\nu)$, obtained using different datasets and approaches, as a function of frequency, following the Eq. \ref{eq:fref} using $F_{cat}(\nu)$ and $F_{app,ref}(\nu)$ as described in Sec. \ref{sec:strong_source}. The frequency dependence of $F_{ref}(\nu)$ comes from the overall bandpass response of the MWA. The uncertainties on each of the measurements were obtained by adding $\Delta F_{ref}(\nu)$ and the 8\% systematic flux density uncertainty of the GLEAM survey for the sources in the declination ($\delta$) range $-72^\circ \leq \delta \leq 18^\circ.5$ \citep{walker2017} in quadrature. It is evident that $F_{ref}(\nu)$ varies smoothly across the MWA observing band. The $F_{ref}(\nu)$ values obtained on 20140928 are systematically higher as compared to other days. We note that the attenuator setting used on this day was higher (14 dB) than what was used on the other three days (10 dB), and we will return to a discussion of these observations later in this section. Until then we focus on the observations made with 10 dB attenuation.

Since calibrator observation of Virgo-A with and without the attenuator was only available for 2014 July 12, and two close-by epochs, 2014 July 11 and 2014 July 13, we have used these epochs to determine $B_{ref}(\nu)$. These calibrator observations were done in contiguous mode at 140--170 and 170--200 $\mathrm{MHz}$. The direction-independent bandpass for these two calibrator observations was estimated following the method described in \cite{sokolowski17}. A mean instrumental gain amplitude has been computed for these observations by averaging over all antennas and both the polarizations. An average instrumental bandpass gain, $B_{tot}(\nu,t_{ref})=\ G_{mean}(t_{ref})B_{inter}(\nu)=\ G_{mean,ref}B_{inter}(\nu)$, spanning the entire MWA band was similarly computed using 20140712 Virgo-A observations, which were done using 10 dB attenuation.  We assume the instrumental gain amplitudes, $G_{mean}(t)$, are similar for the calibrator observations on these close epochs. A ratio of $|G_{mean}|^2$, with and without attenuation, has been computed using the overlapping parts of the band. The ratio has been found to be $r\sim$0.09$\pm$0.003. 
Though the attenuators are calibrated in power units (of the voltage squared), in the MWA signal chain, the attenuation is applied to the analog voltages. 
Hence, for 10 $\mathrm{dB}$ attenuation, we expect a change in $|G_{mean}|^2$ by a factor of 0.1. The ratio $r$ is reassuringly close to the expected value. This suggests that, for the observations done with 14 $\mathrm{dB}$ attenuation, $r$ is likely to lie close to its expected value of $0.04$.
We use the amplitudes of $B_{tot}(\nu,t_{ref})$ obtained using observations on 2014 July 11 and 2014 July 13 and scaled by the factor of $\sqrt{r}$ just determined, as the amplitude of $B_{ref}(\nu)$. It is shown in the right panel of Fig. \ref{fig:fluxscale_polyfit} and reflects the MWA spectral response across the band. 

To make it convenient to use $F_{ref}(\nu)$ and amplitudes of $B_{ref}(\nu)$ for any given frequency, we fit the observed values of $F_{ref}(\nu)$ and the amplitudes of $B_{ref}(\nu)$ with a polynomial.
A fifth-order polynomial is found to provide a good fit:
\begin{equation}
    y = a_5\nu^5+a_4\nu^4+a_3\nu^3+a_2\nu^2+a_1\nu+a_0
    \label{eq:attn}
\end{equation}
where $y$ is either $F_{ref}(\nu)$ or the amplitudes of $B_{ref}(\nu)$.
The best-fit polynomials are shown in Fig. \ref{fig:fluxscale_polyfit} and the polynomial coefficients are listed in Table \ref{tab:poly_coeff}.

All of the MWA solar observations thus far have used one of two attenuation settings discussed here: 10 dB and 14 dB. The observations with 14 dB attenuation used here cover only the part of the MWA band below $\sim133$ $\mathrm{MHz}$. The ratio of the $F_{ref}(\nu)$ values obtained at 10 and 14 dB are remarkably consistent with each other at all frequencies $\lesssim 133$ $\mathrm{MHz}$, with both the mean and median of these numbers being $\sim 1.19$ and the standard deviation being 0.07. To be able to extend the $F_{ref}(\nu)$ for 14 dB to the rest of the MWA band, we take the empirical approach of scaling the best-fit polynomial for $F_{ref}(\nu)$ arrived at for 10 dB observations by a factor of 1.19. Similarly, $|B_{ref}(\nu)|^2$, determined for 10 dB, needs to be multiplied by $0.04/0.1=0.4$ for 14 dB observations.

\begin{table}
    \centering
    \begin{tabular}{|c|c|c|}
    \hline
    Polynomial Coefficient & $F_{ref}(\nu)$ & $amp\ (B_{ref})$\\[0.1cm]
    \hline
    $a_5\times10^{9}$  & $-1.462\pm0.009$ & $-0.7070\pm0.029$\\[0.1cm]
    $a_4\times10^{6}$  & $+1.558\pm0.078$ & $+0.5216\pm0.017$\\[0.1cm]
    $a_3\times10^{4}$  & $-6.250\pm0.024$ & $-1.4825\pm0.005$\\[0.1cm]
    $a_2\times10^{1}$  & $+1.207\pm0.036$ & $+0.2032\pm0.007$\\[0.1cm]
    $a_1\times10^{-1}$  & $-1.159\pm0.026$ & $-0.1350\pm0.005$ \\[0.1cm]
    $a_0\times10^{-2}$  & $+4.719\pm0.076$ & $+0.3449\pm0.015$\\[0.1cm]
    \hline
    \end{tabular}
    \caption{Best-fit Values of Polynomial Coefficients Used in Figure \ref{fig:fluxscale_polyfit}. $\nu$ is in ${\mathrm{MHz}}$.}
    \label{tab:poly_coeff}
\end{table}

Note that, due to inaccuracies in the primary beam model, the observed flux densities can show systematic declination and frequency-dependent biases \citep{sutinjo2015, walker2017}. 
For the purpose of arriving at the best-fit polynomial, we ignore these systematic errors and only take the random errors into account.

\subsection{Variation of instrumental gain in time}
\label{sec:time-variation}
 A database of MWA calibration solutions has recently been made available by \citet{Sokolwski2020}.
This database provides robust amplitude and phase calibration solutions for individual antenna tiles at multiple epochs per day for a large fraction of MWA data available at the data archive hosted by the Pawsey Supercomputing Centre via the MWA ASVO interface\footnote{https://asvo.mwatelescope.org/}.
We use this database to estimate the stability of $B_{inter}(\nu)$ in time and the variation of $G_{mean}(t)$ seen in the MWA data. 
For this exercise, calibration solutions were chosen from this database at intervals of two to three weeks spanning the period from 2013 June to 2020 June, and a normalized bandpass, $B_{inter}(\nu)$, was computed for each epoch. The 1$\sigma$ variation of the amplitude of the $B_{inter}(\nu)$ was found to lie in the range of 4--5\% at edges of the MWA band and 2--3\% in the middle part of the MWA band. The variation in the amplitude of $B_{inter}(\nu)$ is comparable to the $\sim$3\% uncertainty in the $F_{ref}(\nu)$ and that on its best-fit polynomial description. It is also much smaller than the $\sim$8\% uncertainty associated with GLEAM absolute flux density calibration. This implies that the epoch-to-epoch variations in the spectral shape of the $B_{inter}(\nu)$ lead to an insignificant increase in the overall uncertainty in absolute flux density calibration. On the other hand, $G_{mean}(t)$ shows much larger variations of 10--30\% from epoch to epoch. This needs to be corrected in order to avoid leaving a large systematic uncertainty in the absolute flux density estimates.

\subsection{Stability of attenuator response}\label{sec:time-variation-atten}
For the value of $t_{ref}$ we get from Eq. \ref{eq:fref},
\begin{equation}
\begin{split}
F_{ref}(\nu)\ |B_{ref}(\nu)|^2&=\frac{1}{A_{ref}(\nu)}
\end{split}
\end{equation}\label{eq:fref_mult_bref2}
We show the product of $F_{ref}(\nu)$ and $|B_{ref}(\nu)|^2$ in Fig.  \ref{fig:fscale_mult_bref}. As is evident from this figure, it does not show a systematic trend with frequency, and it has a mean and rms of 44 and 4, respectively, for 10 dB attenuation. The scatter observed in this quantity is much smaller than the fractional systematic variations across the band seen in either of $F_{ref}(\nu)$ and $B_{ref}(\nu)$. 
This implies that $A_{ref}(\nu)$ does not have a significant spectral dependence. 
This frequency independence can be made use of for flux density calibration of the solar observations even when calibrator observations in same spectral settings are not available. 
We can use any calibrator observation to determine the un-normalized bandpass gain, $B_{tot}(\nu,t)=G_{mean}(t)\ B_{inter}(\nu, t)$. The stability of the spectral shape of the $B_{inter}(\nu)$, as discussed in Sec. \ref{sec:time-variation}, allows us to take the same $B_{inter}(\nu)$ for both $t$ and $t_{ref}$. We determine the value of the scaling, $S(t)$ ,as:
\begin{equation}
\begin{split}
   S(t) &=\frac{|B_{ref}(\nu)|^2}{|B_{tot}(\nu,t)|^2}\\
   &= \frac{|G_{mean,ref}|^2\ |B_{inter}(\nu)|^2}{|G_{mean}(t)|^2\ |B_{inter}(\nu)|^2}\\
   &= \frac{|G_{mean,ref}|^2}{|G_{mean}(t)|^2} 
\end{split}\label{eq:6}
\end{equation}
$B_{ref}(\nu)$ is needed to obtain the $S(t)$ from any calibrator observation without attenuation in any spectral configuration to scale the $F_{ref}(\nu)$ to $F_{scale}(\nu,\ t)$. 

The consistency of $F_{ref}(\nu)=F_{scale}(\nu,\ t=t_{ref})$ over reference epochs has been demonstrated in left panel of Fig. \ref{fig:fluxscale_polyfit}. $F_{scale}(\nu,\ t)$ has contributions from both attenuator response and instrumental gain. However, it is likely that the consistency of $F_{scale}(\nu,\ t)$ across epochs arises from the individual stability of both the bandpass amplitude or the attenuator response. Thus, one cannot formally claim that this is due to the degeneracy just mentioned. An independent estimate for the stability of the attenuator response can be arrived at by exploring the stability of the bandpass amplitude solutions for these epochs. We have shown the bandpass amplitudes of three epochs, 2014 May 03, July 11, and September 24, in Fig. \ref{fig:epochs_meanband},  when calibrator observations were available. The data quality of 2014 September 27 was poorer for frequencies greater than 150 $\mathrm{MHz}$, and those data are not shown here. These mean bandpass amplitudes show that even data taken months apart are consistent within $\sim$2\%. This is very similar to the variability observed in $F_{scale}(\nu, t)$, and hence implies that the attenuator response must have remained essentially constant across these observations. This is consistent with the expectation that, because they are passive devices, attenuators are not prone to significant evolution in their characteristics, and suggests that it is reasonable to assume that the attenuator performance, $A(t)$,  has remained steady across time. We did not have any calibrator observations with the same frequency range as shown in Fig. \ref{fig:epochs_meanband} for 2020 June 20. Thus, we could not directly compare the bandpass amplitude with the other three epochs. Since the $F_{scale}(\nu,\ t)$ for 2020 June 20 also matched well with other epochs and the time-independent nature of $A$ just mentioned, we have considered 2020 June 20 also as a reference epoch. Since $A$ is independent of time and frequency, we could use the scaling values for $F_{ref}(\nu)$ and $B_{ref}(\nu)$ from 10 dB to 14 dB as mentioned in Sec. \ref{sec:time-variation} for any other epochs. Using Eq. \ref{eq:fref} and scaling values mentioned in Sec. \ref{sec:time-variation}, we find the value of $\frac{1}{A}$ for 14 dB attenuation to be 92.

\subsection{A comparison with GLEAM}
It is instructive to compare the image presented in Section \ref{sec:other_sources} with the typical imaging quality delivered by the MWA GLEAM survey \citep{walker2017}, as it provides a good benchmark for quality of imaging being provided by our solar imaging pipeline (D. Kansabanik et al. 2022, in preparation).
At 108 $\mathrm{MHz}$, using 20140504 data, we obtain an image rms of $\sim 0.72$ $\mathrm{Jy}$ for an integration time of 2 minutes and 2.28 $\mathrm{MHz}$ bandwidth.
The GLEAM survey lists typical rms values at 72 $\mathrm{MHz}$ and 240 $\mathrm{MHz}$ for its integration time of 2 minutes and bandwidth of 7.68 $\mathrm{MHz}$. Using this information, we estimate the rms at 108 $\mathrm{MHz}$ to be about $0.150$ $\mathrm{Jy}$. In addition to the differences in time and frequency integration, an apples-to-apples comparison requires us to also take two other considerations into account. 
The first is the increase in system temperature due to the Sun, and this information is available in \citet{oberoi2017}.
The second consideration is that, because it is an aperture array, the sensitivity of the MWA is a function of the elevation of the pointing direction, which then also needs to be accounted for. 
While the GLEAM observations were carried out at the highest elevation feasible for a given declination strip, the 20140504 observations were carried out at the elevation limit of the MWA. 
Once all these factors are taken into account, it leads to an expected rms of $\sim$0.48 $\mathrm{Jy}$ for a time and frequency integration of 2 $\mathrm{minutes}$ and 2.28 $\mathrm{MHz}$.
This demonstrates that, despite the dynamic range limitations expected due to presence of a very bright and extended source in the FoV, our imaging pipeline is able to achieve an rms noise only $\sim$1.5 times that expected from extrapolation from GLEAM. Details of the calculation of sensitivities are presented in Appendix \ref{app:1}.

\section{Applying the flux scale}\label{sec:apply}
For the MWA solar observations, the following prescription can be used to obtain absolute flux density calibrated images in units of $\mathrm{Jy/beam}$:

\begin{enumerate}
    \item Compute normalized bandpass, $B_{picket}(\nu)$, for each 1.28 $\mathrm{MHz}$ coarse channel for the solar observations independently, and correct for it.
    \item Correct the solar images for the primary beam response using the full embedded element beam model \citep{sokolowski17}  for every 1.28 $\mathrm{MHz}$ coarse channel.
    \item Compute the value of $F_{ref}(\nu)$ and $B_{ref}(\nu)$ corresponding to the value of attenuation used for the observation (10 dB or 14 dB) using the polynomial coefficients in Table \ref{tab:poly_coeff} at the desired observing frequency.
    \item Obtain bandpass gains for any calibrator observation, $B_{tot}(\nu,\ t)$, in any spectral configuration without any additional attenuation from a nearby epoch. Compute $S(t)$ from Eq. \ref{eq:6} using the part of the band overlapping between $B_{tot}(\nu, t)$ and $B_{ref}(\nu)$.
    \item Since $A$ is considered to be independent of time and frequency, $A(\nu,\ t)=A_{ref}(\nu)=A$. From Equations \ref{eq:3}, \ref{eq:fref} and \ref{eq:6}, $F_{scale}(\nu,\ t)$ for any observing epoch, $t$, can be written as
    \begin{equation}
        \begin{split}
            F_{scale}(\nu,\ t)& =\frac{1}{A\ |B_{tot}(\nu,\ t)|^2}\\
            & = \frac{1}{A\ |G_{mean}(t)|^2\ |B_{inter}(\nu)|^2}\\
            & = \frac{S(t)}{A\ |G_{mean,ref}|^2\ |B_{inter}(\nu)|^2}\\
            & = \frac{S(t)}{A\ |B_{ref}(\nu)|^2}\\
            F_{scale}(\nu,t)& = S(t)\ F_{ref}(\nu)
        \end{split}
    \end{equation}
    \item Multiply the primary beam corrected solar images with $F_{scale}(\nu,\ t)$ to obtain the final flux density calibrated image in $\mathrm{Jy/beam}$ units. $F_{scale}(\nu,\ t)$ corrects the MWA bandpass response, temporal variation of the instrumental gain, and the response of the attenuator.
    \item An approximate way to account for the uncertainties due to multiple contributions is to add, in quadrature, an additional 10\% uncertainty to the values obtained from the best-fit polynomial.
\end{enumerate}

This method can also be employed for solar observations with no corresponding calibrator observations, as was the case during early solar observations with the MWA. It has the ability to correct for the large variation in the overall amplitude of the frequency independent gain of the MWA bandpass, which would otherwise be the dominant source of uncertainty, using $S(t)$.
The uncertainty in the $F_{scale}(\nu,t)$ comes primarily from the $\sim$8\% uncertainty of GLEAM flux densities, 3--4\% uncertainty on $F_{ref}(\nu)$ due to the thermal noise, and the 2--5\% variations in the bandshape across epochs. Together, they lead to an overall uncertainty of $\sim$10\% in the final flux density estimates.

\section{Conclusion}\label{sec:conclusion}
We describe a robust flux density calibration method for solar observations with the MWA. 
Not only is the $\sim$10\% uncertainty it delivers is a significant improvement over the  10--60\% uncertainty provided by technique in use thus far \citep{oberoi2017}, the intrinsic simplicity of its application also makes it much less computationally intensive than the earlier approach. 
There are also multiple other significant limitations of the earlier technique that this method overcomes.

The limitations overcome include the requirement of a sufficient number of very short baselines, as well as the dependence of the uncertainty in flux density calibration on the location of the Sun in the sky and the state of solar emission. 
This method can also be applied to any solar observations done with MWA, independent of array configuration, spectral configuration, and attenuator settings. Only two different attenuation settings have been used for MWA solar observations, and the scaling between them has also been determined here. 
The method can also provide flux density calibration for solar data even for the epochs without any matching calibrator observations. This approach has been incorporated in the new version of AIRCARS (D. Kanabanik et al. 2022, in preparation), and will enable routine generation of solar radio images in absolute flux density units.

We find that MWA sensitivity is sufficient to observe some of the stronger flux density calibrators using the attenuation typically used for solar observations even at our highest observing frequency of 240 $\mathrm{MHz}$. This will be a good practice to follow for future solar observations.
Being able to do so will simplify flux density calibration process and will also allow us to take into account any variations in the spectral behavior of $F_{ref}(\nu)$ at scales too fine to be captured by the polynomial fit employed here, if present. This small-scale variations in bandpass amplitudes are evident from Fig. \ref{fig:epochs_meanband}. Understanding and modeling these small-scale variations will be important for characterizing the spectral properties of weak nonthermal emissions like the Weak Impulsive Narrow Band Quiet Sun Emissions \citep[WINQSEs;][S. Mondal et al. 2022, in preparation]{Mondal2020a,Mondal2021}.

This work also demonstrates that the imaging quality delivered by the combination of MWA data and AIRCARS is sufficient to detect multiple background sources in the vicinity of the Sun.
The flux density of the weakest source detected here is $\sim 4.6$ $\mathrm{Jy}$. We detect the closest source at 20 $R_{\odot}$ from the Sun having a flux density of 4.9 $\mathrm{Jy}$. This, combined with the large MWA FoV, is expected to enable many new capabilities/studies that have remained inaccessible until now. For instance, the catalog positions of background sources can be used to improve the astrometric accuracy of solar images. Though one will still have coronal propagation effects to contend with, this will be a step forward with regard to the long-standing issues  that arise because the locations of coronal radio features do not correspond precisely to photospheric or chromospheric features.
This work also suggests that it will be interesting to explore the possibility of conducting interplanetary scintillation (IPS) observations with the Sun in the FoV. 
Currently, IPS observations with the MWA are conducted while keeping the Sun in a null of the MWA beam \citep[e.g.][]{Morgan2018a,Chettri2018}. The large angular size of the Sun and the chromatic nature of the MWA beams, however, severely restrict the observing windows both in time and also frequency, which this approach might help overcome. Combined with wide FoV polarimetric imaging, the ability to see multiple faint background sources can be used to perform image-based polarization leakage correction. This is also an essential step toward developing the ability to measure the Faraday rotation of linearly polarized radiation from background sources due to the magnetized CME plasma, a long-term goal that we are pursuing.

\appendix
\section{Comparison of Image rms Noise with GLEAM survey}\label{app:1}

This appendix presents the calculations of the expected rms noise for the 20140504 observation and its comparison with the GLEAM survey \citep{walker2017}. 
Expected rms noise per polarization, $\Delta F_{X,Y}$, can be written as
\begin{equation}\label{eq:sesitivity}
    \begin{split}
        \Delta F_{X,Y}=\frac{SEFD_{X,Y}}{\sqrt{N_{ant}(N_{ant}-1)\Delta \nu \Delta t}}
    \end{split}
\end{equation}
where $X$ and $Y$ refer to the two orthogonal polarizations, $SEFD$ is the system equivalent flux density, $N_{ant}$ is the number of antennas used for imaging, $\Delta t$ is the total integration time used for imaging, and $\Delta \nu$ the total imaging bandwidth 
\citep{Thomson2017}.
$SEFD$ can also be expressed in terms of effective collecting area, $A_{eff}$, and system temperature, $T_{sys}$, as
\begin{equation}\label{eq:aeff}
    SEFD=\frac{2\ K\ T_{sys}}{A_{eff}},
\end{equation}
where $K$ is the Boltzmann constant
\citep{Thomson2017}.
The value of $A_{eff}$ has been calculated using the Full Embedded Element primary beam model developed by \citet{sokolowski17}.  $T_{sys}$ has the contributions from beam-averaged sky temperature ($T_{sky}$), receiver temperature ($T_{rec}$) and ground pick-up ($T_{pick}$). We use values of $T_{rec}$ and $T_{pick}$ provided by \citet{Daniel2020}.

GLEAM observations were done using the primary beam pointings at higher elevations.
For aperture arrays like the MWA, the sensitivity is a function of elevation and higher elevations offer greater sensitivity.
We choose a GLEAM pointing at 108 $\mathrm{MHz}$ for the same part of the sky as the 20140504 solar observation. 
The theoretical thermal rms noise of GLEAM image, without any contribution from the Sun, was calculated using Eqs. \ref{eq:sesitivity} and \ref{eq:aeff}, as well as the parameters listed in Table \ref{table:sensitivity}. 
The theoretical rms noise for GLEAM was estimated to be $\sim$38 $\mathrm{mJy}$.
The noise obtained in the GLEAM images is $\sim$150 $\mathrm{mJy}$  \citep{walker2017}, $\sim$4 times larger than the theoretical value.

To estimate the theoretical thermal rms for solar images, we add the beam-averaged contribution of the Sun ($T_{\odot}$) to the $T_{sys}$. 
$T_\odot$ at 108 $\mathrm{MHz}$ was estimated to be 159 $\mathrm{K}$ and represents the average of the values at 103 and 117 $\mathrm{MHz}$ from the work of \citet{oberoi2017}, which provides the values for a quiet solar time.
In addition, while the GLEAM survey in general used a full 128 MWA antenna elements (tiles), only 115 of them were used for observations on 20140504. 
Taking these considerations into account leads to a theoretical thermal noise of 120 $\mathrm{mJy}$ for solar observations on 20140504 using the parameters mentioned in Table \ref{table:sensitivity}.
Scaling up the thermal noise by the factor of 4 estimated for GLEAM leads to an expectation of 480 $\mathrm{mJy}$ for the observed rms noise.
The actual value of the rms observed in solar map is 720 $\mathrm{mJy}$, a factor 1.5 higher than the expectations based on GLEAM.

\begin{table*} \label{table:sensitivity}
\centering
    \begin{tabular}{|p{1.7cm}|p{1cm}|p{1cm}|p{1.5cm}|p{1.5cm}|p{1cm}|p{1cm}|p{1cm}|p{1cm}|p{1cm}|p{1cm}|p{1cm}|}
        \hline
        Observation & $A_{eff,X}$\newline$(m^2)$ & $A_{eff,Y}$\newline$(m^2)$ & $SEFD_X$\newline(Jy) & $SEFD_Y$ \newline(Jy) & $T_{sky,X}$\newline(K) & $T_{sky,Y}$\newline(K)  & $T_\odot$\newline(K) & $T_{sys,X}$\newline(K) & $T_{sys,Y}$\newline(K) & $\Delta \nu$\newline(MHz) & $\Delta t$\newline(s) \\[0.1cm]
        \hline
        GLEAM & 17.33 & 13.98 & 134098 & 160703 & 731 & 703 &  0 & 842 & 814 & 7.68 & 120 \\[0.1cm]
        20140504 & 11.61 & 11.69 & 230594 & 226891 & 700 & 691 & 159 & 970 & 961 & 2.28 & 120 \\[0.1cm]
        \hline
    \end{tabular}
    \caption{Parameters used to estimate the theoretical rms noise.}
    * We have used $T_{rec}$ = 91 K and $T_{pick}$ = 20 K.
\end{table*}

\facilities{Murchison Widefield Array (MWA) \citep{lonsdale2009,tingay2013}}

\software{astropy \citep{price2018astropy}, matplotlib \citep{Hunter:2007}, Numpy \citep{Harris2020}, CASA \citep{mcmullin2007}, PyBDSF \citep{mohan2015}, AIRCARS \citep{mondal2019}}

\begin{acknowledgments}
This scientific work makes use of the Murchison Radio-astronomy Observatory (MRO), operated by the Commonwealth Scientific and Industrial Research Organisation (CSIRO).
We acknowledge the Wajarri Yamatji people as the traditional owners of the Observatory site. 
Support for the operation of the MWA is provided by the Australian Government's National Collaborative Research Infrastructure Strategy (NCRIS), under a contract to Curtin University administered by Astronomy Australia Limited. We acknowledge the Pawsey Supercomputing Centre, which is supported by the Western Australian and Australian Governments. 
D.K. and S.M. gratefully acknowledge Barnali Das (NCRA-TIFR) for useful discussions.
We acknowledge Natasha Hurley-Walker and Randall Wayth, both at ICRAR, Curtin University, for considered comments on an earlier draft of the manuscript.
We also gratefully acknowledge the thoughtful and positive comments from the anonymous referee which have helped improve the clarity and the presentation of this work.
D.K., D.O. and A.B. acknowledge support of the Department of Atomic Energy, Government of India, under the project no. 12-R\&D-TFR-5.02-0700. S.M. acknowledges partial support by USA NSF grant AGS-1654382 to the New Jersey Institute of Technology. S.B. is supported by the Hamilton PhD Scholarship at DIAS. We thank the developers of Python 2.7 \citep{van1995python} and the various associated packages, especially Matplotlib \citep{Hunter:2007}, Astropy \citep{price2018astropy} and NumPy \citep{Harris2020}.
This research has also made use of NASA's Astrophysics Data System (ADS).
This research has made use of the NASA/IPAC Extragalactic Database (NED), which is operated by the Jet Propulsion Laboratory, California Institute of Technology, under contract with the National Aeronautics and Space Administration.
\end{acknowledgments}

%%%%%%%%%%%%%%%%%%%%%%%%%%%%%%%%%%%%%%%%%%%%%%%%%%



\bibliographystyle{aasjournal}
\bibliography{bibliography.bib} % if your bibtex file is called example.bib

%%%%%%%%%%%%%%%%%%%%%%%%%%%%%%%%%%%%%%%%%%%%%%%%%%

%%%%%%%%%%%%%%%%% APPENDICES %%%%%%%%%%%%%%%%%%%%%

%%%%%%%%%%%%%%%%%%%%%%%%%%%%%%%%%%%%%%%%%%%%%%%%%%

\label{lastpage}
\end{document}